\documentclass[doublecol]{epl2} 
\usepackage{amsmath}
\usepackage{graphicx}
\usepackage{epsfig}

\title{Localized states at zigzag edges of multilayer graphene 
  and graphite steps}
\shorttitle{Surface states in multilayer graphene and graphite} 

\author{Eduardo V. Castro\inst{1} \and N. M. R. Peres\inst{2} \and
  J. M. B. Lopes dos Santos\inst{1}}
\shortauthor{E. V. Castro \etal}

\institute{
  \inst{1} CFP and Departamento de F\'{\i}sica, Faculdade de Ciências
  Universidade do Porto - P-4169-007 Porto, Portugal\\
  \inst{2} Center of Physics and Departamento de F\'{\i}sica, Universidade
  do Minho - P-4710-057 Braga, Portugal
}

\abstract{We report the existence of zero energy surface states localized at
zigzag edges of $N$-layer graphene. Working within the tight-binding
approximation, and using the simplest nearest-neighbor model, we derive
the analytic solution for the wavefunctions of these peculiar surface
states. It is shown that zero energy edge states in multilayer graphene
can be divided into three families: (i)~states living only on a single
plane, equivalent to surface states in monolayer graphene; (ii)~states
with finite amplitude over the two last, or the two first layers of
the stack, equivalent to surface states in bilayer graphene; 
(iii)~states with finite amplitude over three consecutive layers. Multilayer
graphene edge states are shown to be robust to the inclusion of the
next nearest-neighbor interlayer hopping. We generalize the edge state
solution to the case of graphite steps with zigzag edges, and show
that edge states measured through scanning tunneling microscopy and
spectroscopy of graphite steps belong to family~(i) or~(ii) mentioned
above, depending on the way the top layer is cut.
}

\begin{document}

\maketitle


\section{Introduction}
In the past few years carbon physics presented
new challenges to the scientific community, increasing the list of
rather unusual phenomena occurring in this life support element. On
one hand, the discovery of metal free carbon-based magnetism open
a new research field in fundamental physics, with possible applications
in spin electronics \cite{MPbook06,EK05,EKF07}. On the other, the
isolation of a single graphite layer -- \emph{graphene} -- revealed
an ultra-relativistic system full of unconventional electronic properties,
and regarded with great expectation from the point of view of applications
\cite{GN07,Kts06rev,NGP+rmp07}.

The origin of the observed magnetism in carbon-based materials is
still under debate, but the presence of open edges seem to be an ubiquitous
feature \cite{EKF07}. In proton bombarded graphite, which shows
room temperature ferromagnetism, proton irradiation induces hydrogen-terminated
edges \cite{KE07,OTH+07}. In activated carbon fibers and graphitized
nanodiamond particles -- known as nanographite -- Curie-Weiss behavior
and an enhanced paramagnetic susceptibility has been reported \cite{EK05}.
In these nanographites edges play a predominant role due to the built-in
nano-dimension. Edges are assumed to induce $\pi$-localized electrons
due to surface (edge) states, which has been seen as a key ingredient
to understand carbon's magnetic behavior \cite{EKF07,MPbook06}.
Indeed, the existence of edge states localized at zigzag edges of
single layer graphene, induced either by extended defects or vacancies,
is now well documented and their magnetic behavior has been extensively
reported \cite{PGS+06,WakabBchap06,LFY+04,SCLnat06,EKF07}. 

Despite the positive correlation between edge state magnetism in graphene
single layer and magnetic phenomena in graphite and nanographite,
strictly speaking, neither of them are a single layer of graphene.
Although the interlayer coupling is known to be very small, its effect
is not negligible. To give an example, massless Dirac fermions in
single layer graphene turn out to be massive in bilayer graphene \cite{NGP+rmp07}.
This brings about the question whether edge states are robust to stacking,
or in other words, whether multilayer graphene can support edge states
localized on zigzag edges. Moreover, with the advent of graphene physics,
also graphene multilayers (bilayer, trilayer, ...) were isolated. These
graphene multilayers show interesting properties on their own \cite{GN07,NGP+rmp07},
dissimilar from their single layer constituent, and can be even more
suitable for some device applications \cite{CNM+06,OHL+07,LA08}.
Therefore, the question whether multilayer graphene possesses edge
state physics is of paramount importance.

In this Letter we show that zero energy states localized at zigzag
edges do exist in multilayer graphene. Using the simplest first nearest-neighbor
tight-binding model, we derive the analytical expression for multilayer
graphene edge states and show that their number is always equal to
the number of layers occurring at the edge. The effect of second nearest-neighbor
interlayer hopping is considered, and the robustness of multilayer
graphene edge states is shown. Finally, we generalize the edge state
solution to graphite steps, where experimental evidence for edge states
has been widely reported \cite{EKF07}. The theoretical solution
given in this Letter agrees well with experimental findings. Also,
we predict that edge states in graphite steps should be seen in scanning
tunneling microscopy (STM) even when the step occurs underneath the
first graphite layer.

\begin{figure}[t]
\centerline{
\includegraphics[width=0.9\columnwidth]{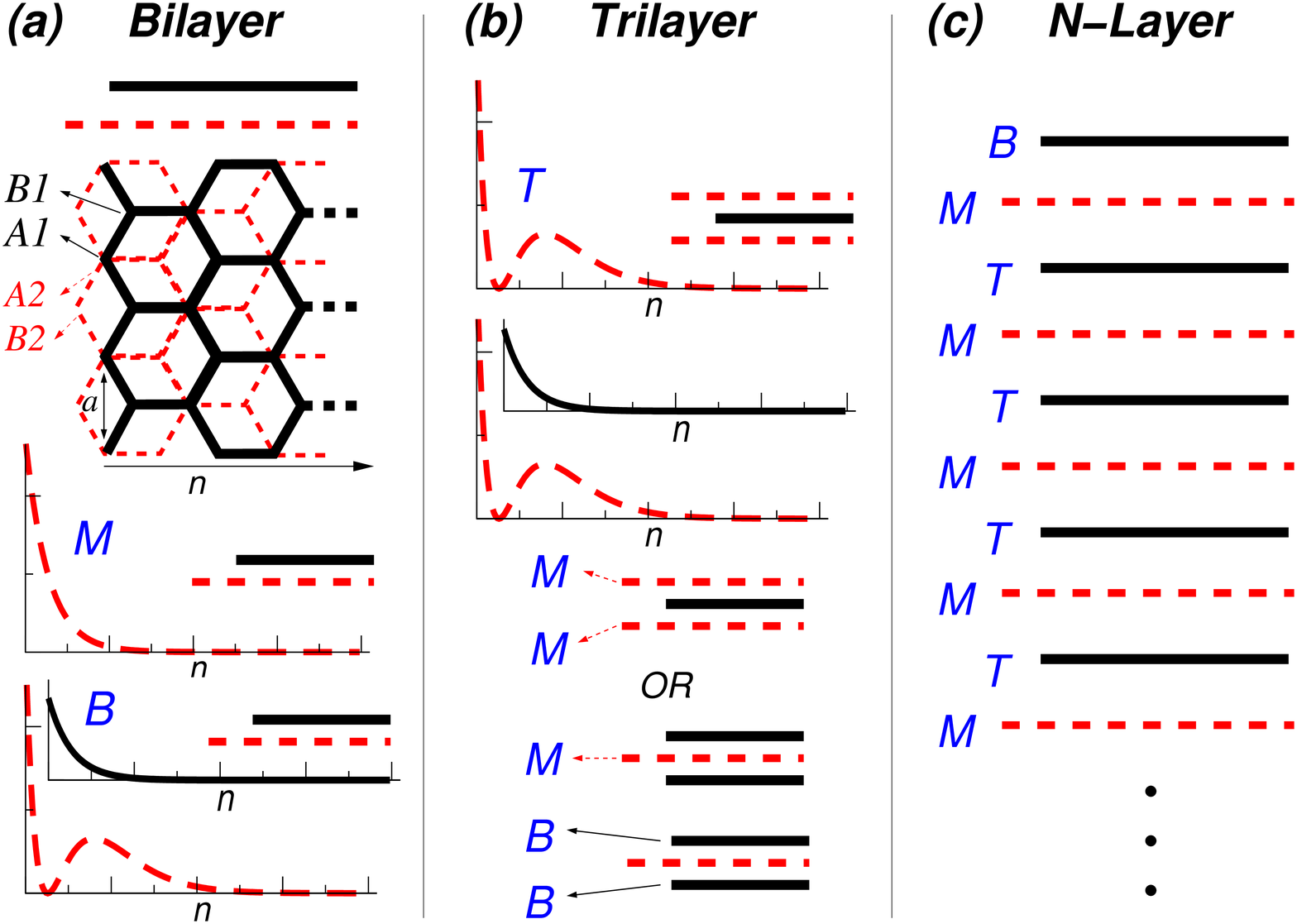}
}

\caption{\label{fig:mgES}(a)~Side and top views of bilayer
graphene, and its two families of edge states: \emph{monolayer} (\emph{M})
and \emph{bilayer} (\emph{B}); the vertical axes represent the associated
charge densities (squared amplitudes). (b)~\emph{Trilayer} family of edge
states (\emph{T}) occurring in trilayer graphene (vertical
axes represent charge densities), and all other possible
edge states (in schematic view). (c)~Edge states in $N$-layer graphene
(in schematic view).}

\end{figure}


\section{Model}
We model $AB$-stacked multilayer graphene as shown
in fig.~\ref{fig:mgES}(a) (for the simplest case of a bilayer),
where non-interacting $\pi$-electrons are allowed to hop only between
$A$ and $B$ sublattices. In what follows we use the terminology
\emph{balcony layers} for layers represented with dashed (red) lines,
and \emph{non-balcony} layers for those represented with full (black)
lines. Without loss of generality we assume all edge atoms belong
to the $A$ sublattice. The zigzag edge breaks translational invariance
along its perpendicular direction, enabling us to write an effective
one-dimensional Hamiltonian for a given momentum $k\in[0,2\pi[$ along
the edge (in units of $a^{-1}$). The first nearest-neighbor tight-binding
Hamiltonian can be written as%
%
\begin{eqnarray}
H_{k} & = & -t\sum_{i}\sum_{n}a_{i;k,n}^{\dagger}(-e^{ik/2}D_{k}b_{i;k,n}+b_{i;k,n-1})\nonumber \\
 &  & -t_{\perp}\sum_{i}^{\bullet}\sum_{n}a_{i;k,n}^{\dagger}b_{i\mp1;k,n}+\textrm{h.c.}\,,\label{eq:Hmgk}\end{eqnarray}
where $a_{i;k,n}$ ($b_{i;k,n}$) is the annihilation operator at
momentum $k$ and position $n$ in sublattice $Ai$ ($Bi$), $i$
is the layer index and $D_{k}=-2\cos(k/2)$. The first term in eq.~(\ref{eq:Hmgk})
describes in-plane hopping while the second term parametrizes the
inter-layer coupling ($t_{\perp}\ll t$). The symbol $\bullet$ indicates
a sum over non-balcony layers. Afterwards we consider the second nearest-neighbor
interlayer hopping between $A$ and $B$ sublattices, which implies
an extra term in eq.~(\ref{eq:Hmgk}) given by $-\gamma_{3}\sum_{i}^{\bullet}\sum_{n}b_{i;k,n}^{\dagger}(e^{-ik}a_{i\mp1;k,n}-e^{-ik/2}D_{k}a_{i\mp1;k,n+1})+\textrm{h.c.}$,
where $\gamma_{3}\sim t_{\perp}\ll t$.


\section{Edge states in $N$-layer graphene}
Multilayer graphene edge
states are investigated by solving the Schr\"odinger equation, $H_{k}\left|\psi_{k}\right\rangle =E_{k}\left|\psi_{k}\right\rangle $.
The wavefunction $\left|\psi_{k}\right\rangle $ is written as a linear
combination of the site amplitudes along the edge's perpendicular
direction, $\left|\psi_{k}\right\rangle =\sum_{n}\sum_{i}[\alpha_{i}(k,n)\left|a_{i},k,n\right\rangle +\beta_{i}(k,n)\left|b_{i},k,n\right\rangle ],$
where we have introduced the one-particle states $\left|c_{i},k,n\right\rangle =c_{i;k,n}^{\dagger}\left|0\right\rangle $,
with $c_{i}=a_{i},b_{i}$. In addition we require the boundary conditions
$\alpha_{i}(k,n\rightarrow\infty)=\alpha_{i}(k,-1)=\beta_{i}(k,n\rightarrow\infty)=\beta_{i}(k,-1)=0$,
accounting for the existence of the edge at $n=0$. Within our model,
the Fermi energy of multilayer graphene always occurs at zero energy.
Therefore, we expect zero energy edge states to have interesting physical
consequences, and we set $E_{k}=0$. As a result, the two sublattices
become completely decoupled, and only the sublattice to which edge
atoms belong can support edge states.\footnote{In the ribbon geometry the two sublattices are equivalent, 
supporting edge states localized in opposite ribbon edges. In the 
semi-infinite system, only those localized in the edge 
sublattice survive \cite{WFA+99,CPL+07}.} This means that
we always have $\beta_{i}(k,n)=0$.

It was recently shown \cite{CPL+07} that bilayer graphene supports
two types of zero energy edge states localized at zigzag edges for
$2\pi/3<k<4\pi/3$: one type restricted to the balcony layer and coined
\emph{monolayer family}, with amplitudes equivalent to edge states
in single layer graphene,%
%
\begin{equation}
\alpha_{2}(k,n)=\alpha_{2}(k,0)D_{k}^{n}e^{-i\frac{k}{2}n}\,;\label{eq:SSmf}\end{equation}
and a new type coined \emph{bilayer family}, with finite amplitudes
over the two layers,%
%
\begin{eqnarray}
\alpha_{1}(k,n) & = & \alpha_{1}(k,0)D_{k}^{n}e^{-i\frac{k}{2}n}\,,\nonumber \\
\alpha_{2}(k,n) & = & -\alpha_{1}(k,0)D_{k}^{n-1}\frac{t_{\perp}}{t}e^{-i\frac{k}{2}(n-1)}\Big(n-\frac{D_{k}^{2}}{1-D_{k}^{2}}\Big)\,,\nonumber \\
 &  & \,\,\,\,\label{eq:SSbf}\end{eqnarray}
where the normalization constants are given by $|\alpha_{2}(k,0)|^{2}=1-D_{k}^{2}$
and $|\alpha_{1}(k,0)|^{2}=(1-D_{k}^{2})^{3}/[(1-D_{k}^{2})^{2}+t_{\perp}^{2}/t^{2}]$.
The charge densities (squared amplitudes) associated with
the two families of edge states are represented in fig.~\ref{fig:mgES}(a).
Let us now consider a trilayer as shown in fig.~\ref{fig:mgES}(b),
where a non-balcony layer is sandwiched between two balcony layers.
Clearly, the bilayer family is not an edge state solution for this
trilayer, as any finite amplitude at a non-balcony layer implies,
through eq.~(\ref{eq:Hmgk}), a finite amplitude over adjacent layers.
We note, however, that our model ignores the coupling between next
nearest-layers.\footnote{This is a reasonable approximation since in the 
Slonczewski-Weiss-McClure parametrization 
$\gamma_{2},\gamma_{5}\ll t_{\perp},\gamma_{3}$ \cite{NGP+rmp07}.} 
Thus, if we construct a trilayer wavefunction
whose amplitudes over balcony/non-balcony layers mimic those for the
bilayer family, it is guaranteed, apart from a normalization factor,
that we have an edge state solution. More precisely, we arrive at
a new type of edge state with finite amplitudes over three consecutive
layers -- \emph{trilayer family} -- whose analytic form can be written
as\begin{eqnarray}
\alpha_{1}(k,n) & = & -\alpha_{2}(k,0)D_{k}^{n-1}\frac{t_{\perp}}{t}e^{-i\frac{k}{2}(n-1)}\Big(n-\frac{D_{k}^{2}}{1-D_{k}^{2}}\Big)\,,\nonumber \\
\alpha_{2}(k,n) & = & \alpha_{2}(k,0)D_{k}^{n}e^{-i\frac{k}{2}n}\,,\nonumber \\
\alpha_{3}(k,n) & = & -\alpha_{2}(k,0)D_{k}^{n-1}\frac{t_{\perp}}{t}e^{-i2)(n-1)}\Big(n-\frac{D_{k}^{2}}{1-D_{k}^{2}}\Big)\,,\nonumber \\
 &  & \,\,\,\,\,\,\label{eq:SStf}\end{eqnarray}
where the normalization constant is given by $|\alpha_{2}(k,0)|^{2}=(1-D_{k}^{2})^{3}/[(1-D_{k}^{2})^{2}+2t_{\perp}^{2}/t^{2}]$.
The charge density (squared amplitude) associated with
the trilayer family of edge states is represented in fig.~\ref{fig:mgES}(b). 

Additionally, the trilayer we have been discussing also supports edge
states of the monolayer family localized at balcony layers, as schematically
shown in fig.~\ref{fig:mgES}(b). In fact, this is a general result.
Balcony layers have an edge sublattice which is not connected through
$t_{\perp}$ to adjacent layers. Thus, the monolayer family is always
an edge state solution in $N$-layer graphene. Even more generally,
we can look at a balcony layer as a buffer layer. As can be seen from
eq.~\eqref{eq:Hmgk}, a finite amplitude over a balcony layer does
not imply finite amplitudes over adjacent layers. For the trilayer
shown at the bottom of fig.~\ref{fig:mgES}(b), where a balcony layer
is sandwiched between two non-balcony layers, monolayer edge states
certainly exist at the middle balcony layer. But because of the buffer
layer character, also bilayer edge states are present, localized either
at the two top or the two bottom layers. An immediate consequence
of the buffer layer concept is the fact that the trilayer family is
the most general edge state family we can have, and exists localized
at any non-balcony layer and its two adjacent layers, with all other
site amplitudes equal to zero. Therefore, we have three families of
edge states occurring in multilayer graphene: (i) monolayer family
for each balcony layer, eq.~\eqref{eq:SSmf}; (ii) bilayer family
for each non-balcony layer that starts and/or ends the multilayer,
eq.~\eqref{eq:SSbf}; (iii)\emph{ }trilayer family for each non-balcony
layer sandwiched between two balcony ones, eq.~\eqref{eq:SStf}.
This is schematically represented in fig.~\ref{fig:mgES}(c). Note
that the number of edge state families is always equal to the number
of edge layers.


\section{Effect of $\gamma_{3}$}
\begin{figure}[t]
\centerline{
\includegraphics[width=0.9\columnwidth]{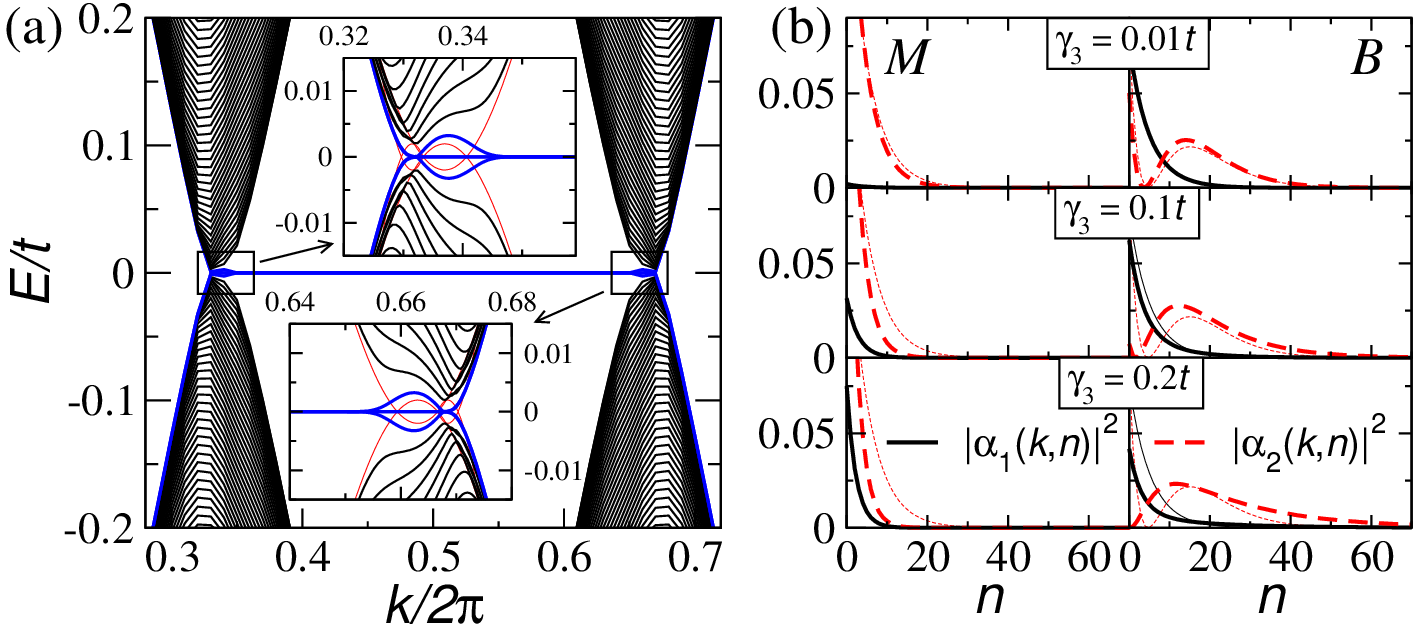}
}

\caption{\label{fig:blg3}(a)~Energy spectrum for a bilayer
ribbon with zigzag edges at finite $\gamma_{3}$. (b)~Charge density
for bilayer graphene edge states, \emph{monolayer} (\emph{M}) and
\emph{bilayer} (\emph{B}) families, at $k/2\pi=0.35$. Thin lines
show the $\gamma_{3}=0$ result.}

\end{figure}

In multilayer graphene the effect
of $\gamma_{3}$ is of the order of $t_{\perp}$, and should be included
in a consistent edge state solution. The buffer layer concept introduced
previously, however, does not survive at a finite $\gamma_{3}$. In
order to generalize the edge state solution to the present case we
use the transfer matrix technique, following ref.~\cite{CPL+07}.
For bilayer graphene the transfer matrix, defined as $[\alpha_{1}(k,n),\alpha_{2}(k,n)]^{T}=e^{-ikn/2}\mathbf{T}(2)^{n}[\alpha_{1}(k,0),\alpha_{2}(k,0)]^{T}$,
is given by\begin{equation}
\mathbf{T}(2)=\left[\begin{array}{cc}
u & v\\
x & D_{k}\end{array}\right]\,,\label{eq:2Lg3TM}\end{equation}
where $u=D_{k}(1-\xi)$, $v=-\frac{\gamma_{3}}{t}e^{-ik/2}(1-D_{k}^{2})$,
and $x=-\frac{t_{\perp}}{t}e^{ik/2}$, with $\xi=t_{\perp}\gamma_{3}/t^{2}$.
The edge states are completely determined by the eigenvalues $\lambda_{\pm}$
and eigenvectors $\chi^{\pm}$ of the transfer matrix. If $|\lambda_{\pm}|<1$,
then edge states exist and are given by $[\alpha_{1}(k,n),\alpha_{2}(k,n)]^{T}\propto e^{-ikn/2}\lambda_{\pm}^{n}\chi^{\pm}$,
apart from a normalization constant. Diagonalizing eq.~\eqref{eq:2Lg3TM}
we obtain $\lambda_{\pm}=D_{k}(1-\xi/2)\pm\sqrt{\xi}\sqrt{D_{k}^{2}(\xi/4-1)+1}$.
Simple algebra shows that for $\lambda_{+}$ the convergence condition
implies $2\cos^{-1}(\sqrt{1+\xi}/2)<k<2\cos^{-1}[-(1-\xi)/2]$ or
$4\pi/3<k<2\cos^{-1}(-\sqrt{1+\xi}/2)$, while for $\lambda_{-}$
it implies $2\cos^{-1}(\sqrt{1+\xi}/2)<k<2\pi/3$ or $2\cos^{-1}[(1-\xi)/2]<k<2\cos^{-1}(-\sqrt{1+\xi}/2)$.
We conclude that bilayer graphene still has two families of edge states
for $\gamma_{3}\neq0$, though the $k$ range is slightly changed
when compared with the $\gamma_{3}=0$ case. In particular, we have
only one family for $k\in[2\pi/3,2\cos^{-1}[(1-\xi)/2]]$ and $k\in[2\cos^{-1}[-(1-\xi)/2],4\pi/3]$,
although the existence of edge states for $k<2\pi/3$ and $k>4\pi/3$
compensates this reduction, and we still have edge states for $1/3$
of the possible $k$'s, as in the $\gamma_{3}=0$. As a test to what
has just been said, we have numerically computed the energy spectrum
for a bilayer ribbon with zigzag edges ($t_{\perp}=\gamma_{3}=0.2t$
and width $400$ unit cells). The result is shown in fig.~\ref{fig:blg3}(a).
Four flat bands at zero energy are clearly seen, and can be identified
with the abovementioned two families of edge states, two per edge.
The insets reveal the $k$ restrictions mentioned before. In fact,
the values of $k$ that limit the existence or number of edge states
coincide with the Dirac points and satellite Fermi points that arise
when $\gamma_{3}\neq0$ \cite{MF06}, as indicated by the thin red
lines (the mismatch is due to the finite width of the ribbon, and
consequent edge state overlap). The transfer matrix eigenvectors can
be written as $\chi^{\pm}=[\lambda_{\pm}-D_{k},-\frac{t_{\perp}}{t}e^{ik/2}]^{T}$,
from which we can write two families of wavefunctions, $\left|\psi_{\pm}\right\rangle =C_{\pm}\sum_{n=0}^{\infty}e^{-ikn/2}\lambda_{\pm}^{n}\chi^{\pm}$,
where the normalization constant is given by $C_{\pm}=[(1-|\lambda_{\pm}|^{2})/(|\lambda_{\pm}-D_{k}|^{2}+t_{\perp}^{2}/t^{2})]^{1/2}$.
Note, however, that $\chi^{+}$ and $\chi^{-}$ are not orthogonal,
implying the non-orthogonality of the two solutions $\left|\psi_{\pm}\right\rangle $.
It is convenient to orthogonalize $\left|\psi_{-}\right\rangle $
with respect to $\left|\psi_{+}\right\rangle $, whose result can
be written as $|\tilde{\psi}_{-}\rangle=(|\psi_{-}\rangle-\langle\psi_{+}|\psi_{-}\rangle|\psi_{+}\rangle)/(1-|\langle\psi_{+}|\psi_{-}\rangle|^{2})$,
where $\langle\psi_{+}|\psi_{-}\rangle=C_{+}C_{-}(t_{\perp}^{2}/t+\xi D_{k}^{2}-\xi)/(1-D_{k}^{2}+\xi)$.
In fig.~\ref{fig:blg3}(b) we show the squared amplitudes associated
with $|\psi_{+}\rangle$ (left) and $|\tilde{\psi}_{-}\rangle$ (right).
The thin lines represent the edge states in bilayer graphene for $\gamma_{3}=0$,
as given by eqs.~\eqref{eq:SSmf} and~\eqref{eq:SSbf}. Clearly,
as long as $\gamma_{3}\ll t_{\perp}$, we can identify $|\psi_{+}\rangle$
with the \emph{monolayer} family and $|\tilde{\psi}_{-}\rangle$ with
the \emph{bilayer} family discussed previously. For $\gamma_{3}\sim t_{\perp}$,
the edge state $|\psi_{+}\rangle$, former monolayer family, already
has an appreciable weight on both layers.

\begin{figure}[t]
\centerline{
\includegraphics[width=0.98\columnwidth]{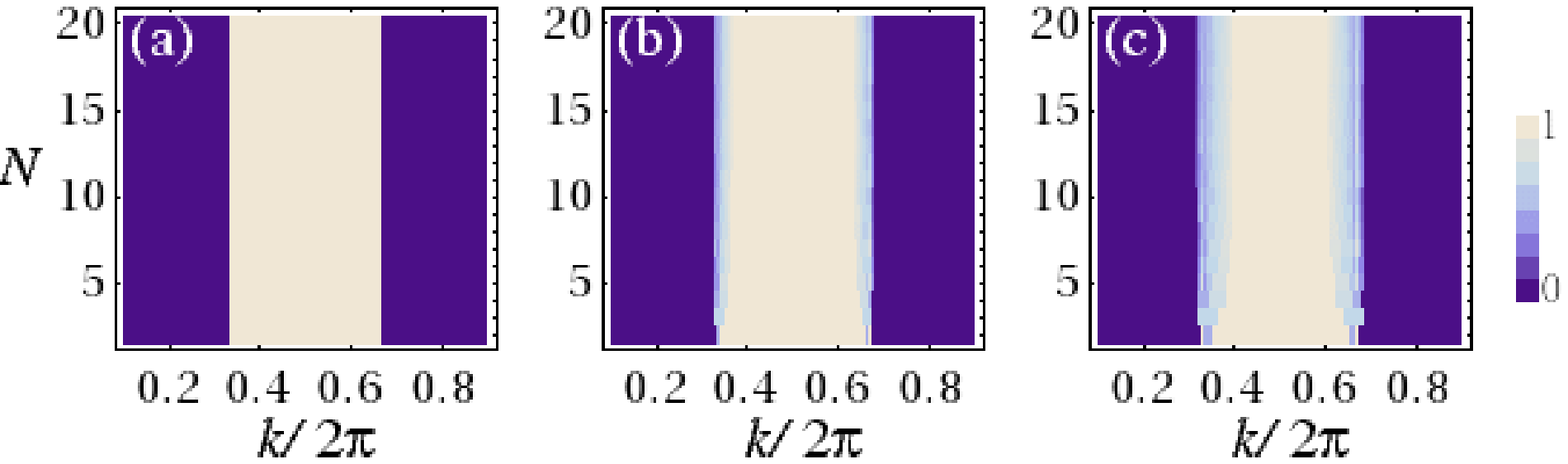}
}

\caption{\label{fig:mlSSg3}Number of edge states per layer
in $N$-layer graphene as a function of $k$: (a)~$\gamma_{3}=0$;
(b)~$\gamma_{3}=t_{\perp}$; (c)~$\gamma_{3}=2t_{\perp}$. We set
$t_{\perp}=0.2t$.}

\end{figure}

The analysis made for bilayer graphene with $\gamma_{3}\neq0$ can
be extended to $N$-layer graphene. Defining the transfer matrix as
$[\alpha_{1}(k,n),\dots,\alpha_{N}(k,n)]^{T}=e^{-ikn/2}\mathbf{T}(N)^{n}[\alpha_{1}(k,0),\dots,\alpha_{N}(k,0)]^{T}$,
we can derive its general pattern for a given number of layers $N$,\begin{equation}
\mathbf{T}(N)=\left[\begin{array}{cccccc}
u & v & -\xi D_{k}\\
x & D_{k} & x\\
-\xi D_{k} & v & u & v & -\xi D_{k}\\
 &  & x & D_{k} & x\\
 &  & -\xi D_{k} & v & u & \cdots\\
 &  &  &  & \vdots & \ddots\end{array}\right]\,.\label{eq:NLg3TM}\end{equation}
The periodic structure of $\mathbf{T}(N)$ is readily identified,
and the transfer matrix for any multilayer graphene is easily constructed.
By diagonalizing eq.~\eqref{eq:NLg3TM}, and checking whether the
eigenvalues $\lambda$ satisfy $|\lambda|<1$, we can conclude about
the existence of edge states in $N$-layer graphene for $\gamma_{3}\neq0$.
In fig.~\ref{fig:mlSSg3} the number of edge states per layer is
shown in the plane \emph{number of layers} ($N$) vs $k$. Panel~\ref{fig:mlSSg3}(a)
confirms what has been said for $\gamma_{3}=0$: same number of edge
states as the number of layers for $2\pi/3<k<4\pi/3$. As shown in
panels~\ref{fig:mlSSg3}(b) and~\ref{fig:mlSSg3}(c), around the
Dirac points the number of edge states for $\gamma_{3}\neq0$ may
be smaller than the number of layers, as we have seen for $N=2$.
However, there is always a broad region in between the Dirac points
where the number of edge states equals the number of layers. Thus,
we conclude that multilayer graphene edges states are robust to second
nearest-neighbor interlayer hopping. This result agrees with first
principles calculations, where the presence of edge states and edge
magnetism have been reported in bilayer \cite{LSP+05,SMD+08} and
graphite \cite{MNF99} zigzag nanoribbons.

\section{Graphite steps}
\begin{figure}[t]
\begin{center}
\includegraphics[width=0.9\columnwidth]{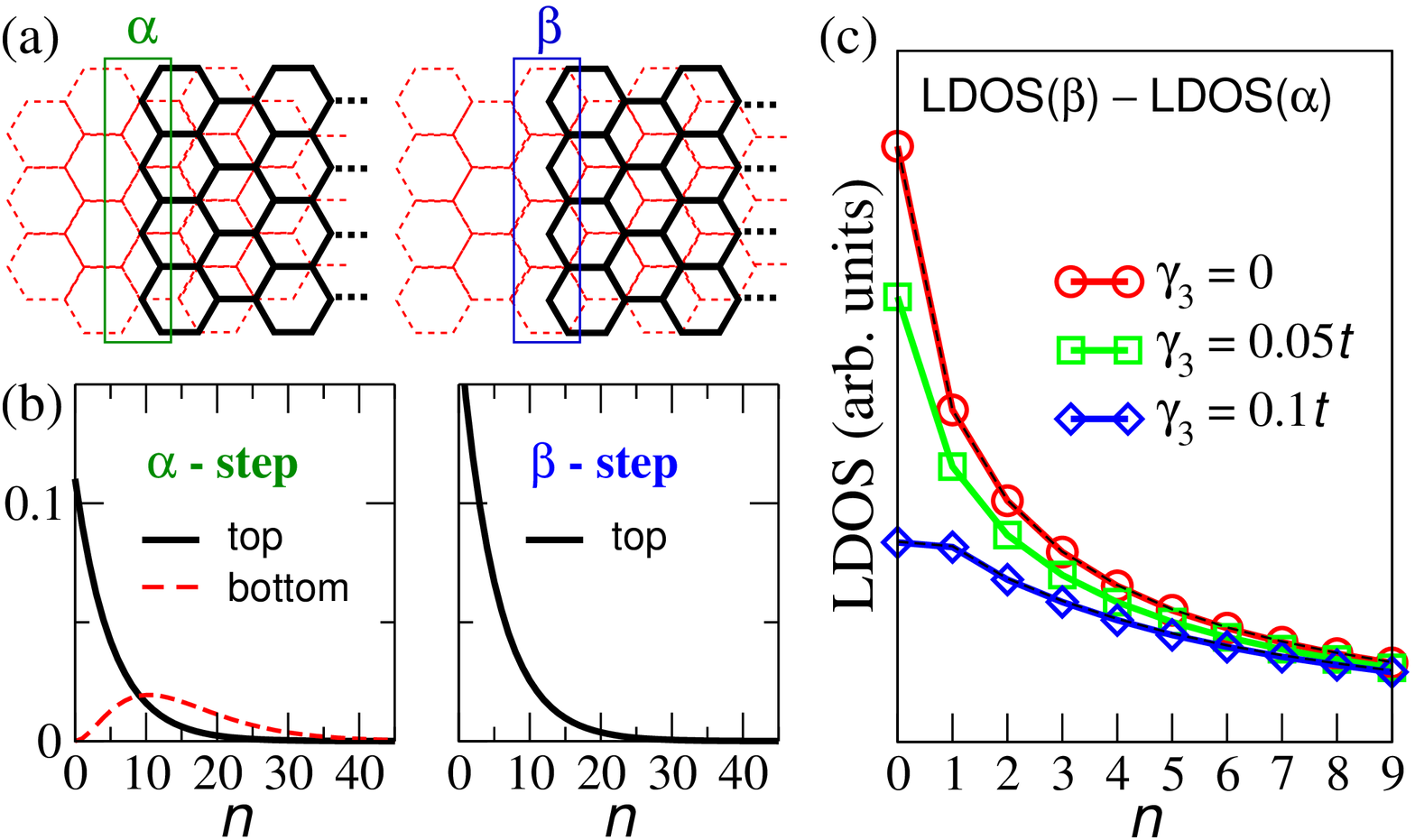}\\
\includegraphics[width=0.9\columnwidth]{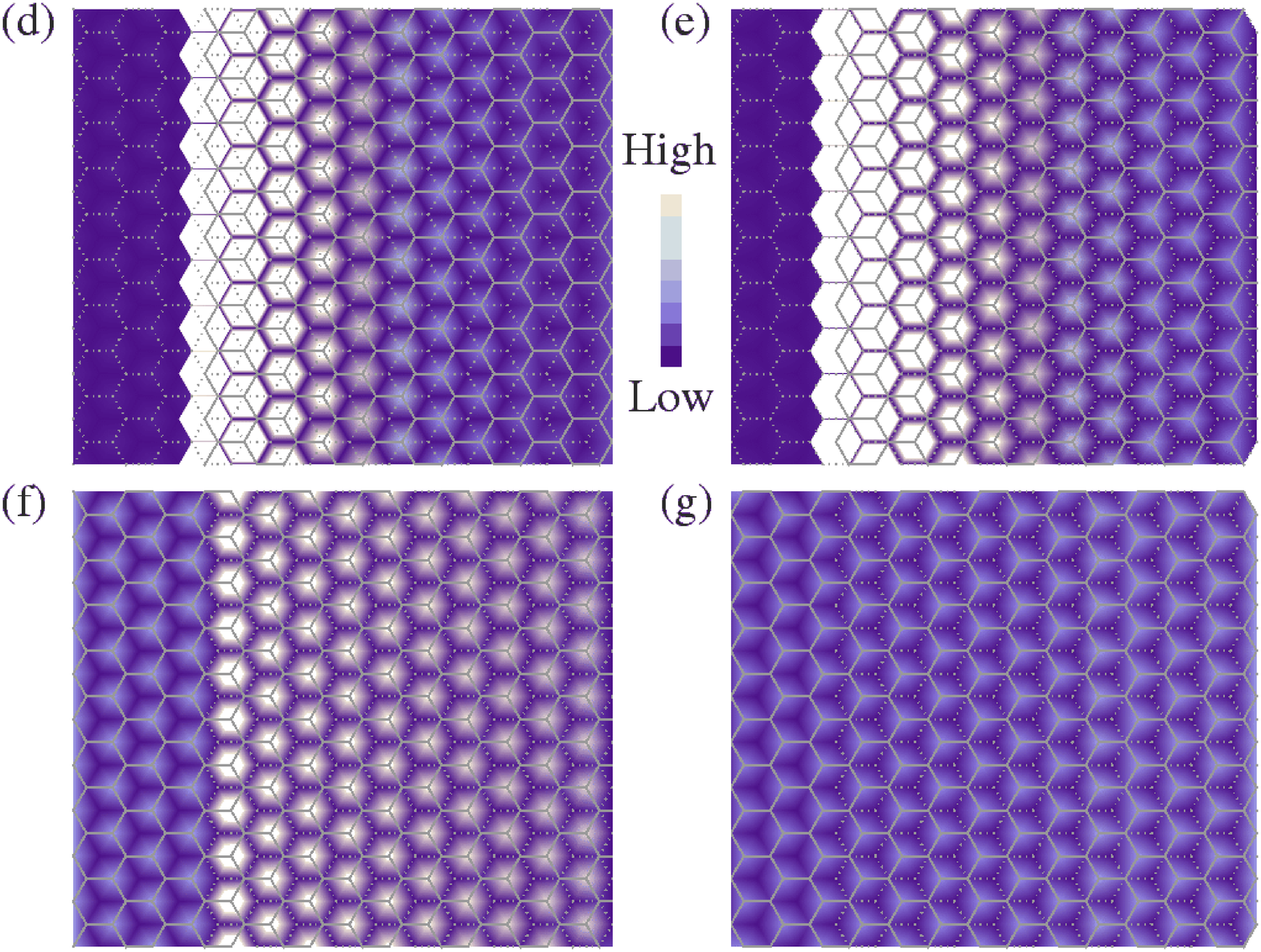}
\end{center}

\caption{\label{fig:step}(a)~Possible zigzag steps on graphite's
surface. (b)~Charge density for edge states at $\alpha$-type and
$\beta$-type step-edges ($ka/2\pi=0.35$). (c)~LDOS difference between
$\beta$- and $\alpha$-type steps as a function of $n$. (d)-(e)~Top
layer LDOS map for $\alpha$- and $\beta$-steps, respectively. (f)-(g)~Underlying
layer LDOS map for $\alpha$- and $\beta$-steps, respectively. We
set $t_{\perp}=0.1t$.}

\end{figure}

Finally, we generalize the edge state solution
to graphite steps, where experimental evidence for edge states has
been widely reported \cite{EKF07,KWD+00,Niim,KFE+05,NMK+06,KFE+06,banerjee2006,SSS+06}.
The local density of states (LDOS) peak seen in STM of graphite steps
has been interpreted as the experimental confirmation of the theoretically
predicted single layer edge states \cite{japonese}. However, we
can easily convince ourselves that monolayer edge states do not always
provide an eigenstate for a zigzag step-edge. To see why, we consider
fig.~\ref{fig:step}(a), where the two possible zigzag step-edges
on the surface of graphite are shown. These two terminations are denoted
$\alpha$-type and $\beta$-type. For an $\alpha$-type termination
the edge carbon atoms occur exactly on top of carbon atoms of the
underlying layer, while for a $\beta$-type termination they occur
at the center of the hexagons. Obviously, monolayer edge states as
given by eq.~\eqref{eq:SSmf} cannot be eigenstates for $\alpha$-steps,
as some finite amplitude must be induced on the second layer through
eq.~\eqref{eq:Hmgk}. For $\beta$-steps, however, single layer edge
states are indeed eigenstates.

In order to have a step-edge we need at least two graphene layers.
Indeed, step-edges are easily obtained from bilayer graphene just
by growing one of the layers beyond the edge. So, localized states
at graphite steps can be understood by studying generalized bilayers
where bottom and top graphene layers have different widths. The two
families of edge states we have found to exist at zigzag edges of
bilayer graphene remain eigenstates even when one of the layers is
wider than the other. In particular, the solution given by eq.~\eqref{eq:SSmf}
and the non-orthogonalized solution for edge states of the bilayer
family given by %
%
\begin{eqnarray}
\alpha_{1}(k,n) & = & \alpha_{1}(k,0)D_{k}^{n}e^{-i\frac{k}{2}n},\nonumber \\
\alpha_{2}(k,n) & = & -\alpha_{1}(k,0)nD_{k}^{n-1}\frac{t_{\perp}}{t}e^{-i\frac{k}{2}(n-1)}\,,\label{eq:SSbfNorth}\end{eqnarray}
{[}a simple linear combination of eqs.~\eqref{eq:SSmf} and~\eqref{eq:SSbf}],
can be adapted to the generalized bilayer just by adjusting the unit
cell index~$n$. Then, Gram-Schmidt orthogonalization gives the final
solution. We should note, however, that the overlap between the two
types of edge states is exponentially suppressed as the difference
in layer width gets larger. When the width of one layer becomes infinite
-- the case of a perfect step -- only one of the two possible solutions
exists. Therefore, the possible localized solutions for zigzag step-edges
(occurring at $n=0$) are:\begin{eqnarray}
\alpha_{\textrm{top}}(k,n) & = & C_{k}D_{k}^{n}e^{-i\frac{k}{2}n}\,,\nonumber \\
\alpha_{\textrm{bottom}}(k,n) & = & -C_{k}nD_{k}^{n-1}\frac{t_{\perp}}{t}e^{-i\frac{k}{2}(n-1)}\,,\label{eq:alphaStep}\end{eqnarray}
for an $\alpha$-type step, and\begin{equation}
\alpha_{\textrm{top}}(k,n)=(1-D_{k}^{2})D_{k}^{n}e^{-i\frac{k}{2}n}\,,\label{eq:betaStep}\end{equation}
for a $\beta$-type step, where $n\geq0$ and the normalization constant
in eq.~\eqref{eq:alphaStep} is given by $|C_{k}|^{2}=(1-D_{k}^{2})^{3}/[(1-D_{k}^{2})^{2}+(1+D_{k}^{2})t_{\perp}^{2}/t^{2}]$.
As in edge states discussed previously, the amplitudes in eqs.~(\ref{eq:alphaStep})
and~(\ref{eq:betaStep}) refer to sites belonging to the same sublattice
as the edge carbon atoms, while the amplitudes at the other sublattice
are zero. Example charge densities for the given edge states are shown
in fig.~\ref{fig:step}(b) for both the $\alpha$-type and the $\beta$-type
step-edges.

As can be seen from eqs.~\eqref{eq:alphaStep} and~\eqref{eq:betaStep},
or by inspection of fig.~\ref{fig:step}(b), there is an apparent
asymmetry between the two families of edge states: edge states at
$\beta$-type steps live only on the top layer, while at $\alpha$-type
steps both the top layer and the underlying layer have a finite edge
state amplitude. Consequently, we expect a similar asymmetry to be
present in the LDOS peak induced by edge states at the Fermi level,
which, ultimately, should be seen with STM. To better appreciate this
effect, we have computed the LDOS of a generalized bilayer -- bottom
layer wider than the top layer -- using the recursive Green's function
method \cite{Hayd80}. The calculated LDOS, which was accumulated
in the range $0.01t$ near the Fermi energy, should be proportional,
in the simplest approximation, to the local tunnel currents in the
experimental STM images \cite{TH85,NMK+06}. In fig.~\ref{fig:step}(c)
we show, for the top layer (edge sublattice), the LDOS difference
between $\beta$-type and $\alpha$-type terminations as we move away
from the step at $n=0$. As expected, the LDOS at $\beta$-steps is
higher and extends further into the bulk, a trend that is still present
for realistic values of $\gamma_{3}$, as shown in fig.~\ref{fig:step}(c).
This behavior agrees with STM results, where two types of edge states
with different penetration depths have been seen \cite{KFE+06}.
Edge states with reduced penetration depth have been observed at $\alpha$-type
steps, whereas at $\beta$-type steps the edge states extend further
into the bulk, as we have obtained here for the top layer component.
However, right at the edge, the STM intensity has been found to be
higher at $\alpha$-steps than $\beta$-steps \cite{KFE+06}. According
to our analytical result the opposite should be seen. This discrepancy
is most probably due to edge state admixture, as experimentally both
$\alpha$-type and $\beta$-type steps coexist on the same step-edge. 

In fig.~\ref{fig:step}(d) and \ref{fig:step}(e) we show the top
layer LDOS map for $\alpha$- and $\beta$-steps, respectively. The
former presents not only a reduced penetration depth, as previously
discussed, but also higher intensity at sites connected to the underlying
layer through $t_{\perp}$, as opposed to standard LDOS maps on the
surface of bilayer graphene and graphite. This behavior is characteristic
of edge states at $\alpha$-type steps, as given by eq.~\eqref{eq:alphaStep}.
fig.~\ref{fig:step}(f) and \ref{fig:step}(g) show the underlying
layer LDOS map for $\alpha$- and $\beta$-steps, respectively. As
edge states at $\alpha$-steps {[}eq.~\eqref{eq:alphaStep}] have
a finite amplitude over the underlying layer, the LDOS map for this
layer shows an increased intensity at and near the step {[}fig.~\ref{fig:step}(f)],
although the lattice discontinuity only exists at the other layer.
As a consequence, we expect $\alpha$-steps to be detected in STM
experiments even when they occur underneath the top layer. This feature
is not seen in $\beta$-type steps {[}fig.~\ref{fig:step}(g)].


\section{Conclusions}
We have demonstrated the existence of zero energy
states localized at zigzag edges of multilayer graphene and graphite
steps. Stability to the presence of interlayer hopping $\gamma_{3}$
has been shown. The electron-hole symmetry breaking terms $\gamma_{4}$
(interlayer) and $t'$ (inplane) are expected to induce edge state
dispersion, but not to qualitatively modify the present results \cite{PGN06,SMS06}.
It should be noted that only perfect zigzag edges have been discussed
here. However, we expect edge state properties to be present in multilayer
graphene and graphite steps even for irregular edges, as long as some
zigzag units are present, as recently demonstrated for single layer
graphene \cite{KH08,BS08}. On the other hand, zigzag edges have
been recently observed in epitaxial graphene monolayer \cite{PCB+08},
providing the first indication that edge shape can be a controllable
parameter in the future. Our findings are relevant in the context
of carbon based magnetism, where edge states seem to play an important
role \cite{EKF07,MPbook06}, and also in the context of graphene
physics, where the reported self-doping in monolayer graphene \cite{PCB+08}
and suppression of conductance fluctuations near the neutrality point
in bilayer and trilayer graphene \cite{SPL07} can be seen as edge
states driven effects.


\acknowledgments

E.V.C., N.M.R.P., and J.M.B.L.S. acknowledge financial support from
POCI 2010 via project PTDC/FIS/64404/2006. 


\end{document}